\documentclass[5p,sort&compress]{elsarticle}
\usepackage{amssymb}
\usepackage{amsmath}
\usepackage{units}
\usepackage{slashed}
\usepackage{graphicx}
\usepackage{hyperref}

\renewcommand{\l}{\left(}
\renewcommand{\r}{\right)}

\newcommand{\GNN}{\Gamma_{\chi\to N_1N_1}}
\newcommand{\eV}{\text{\,eV}}
\newcommand{\keV}{\text{\,keV}}
\newcommand{\MeV}{\text{\,MeV}}
\newcommand{\GeV}{\text{\,GeV}}
\newcommand{\cL}{\mathcal{L}}
\newcommand{\cH}{\mathcal{H}}

\begin{document}

%%%%%%%%%%%%%%%%%%%%%%%%%%%%%%%%%%%%%%%%%%%%%%%%%%%%%%%%%%%%%%%%%%%%%%%%
%\title{Creating the 7 keV Dark Matter with the Inflaton}
%\title{Light Inflaton Producing 7 keV Dark Matter Sterile Neutrino\\ 
%or \\
\title{Relic Gravity Waves and 7\keV{} Dark Matter from a GeV scale
  inflaton}
%(showing up (to be checked) in B-physics)} 

\author[cern,uc,bnl]{F.~L.~Bezrukov}
\ead{Fedor.Bezrukov@uconn.edu}
\author[inr,mipt]{D.~S.~Gorbunov}
\ead{gorby@ms2.inr.ac.ru}

\address[cern]{CERN, CH-1211 Gen\`eve 23, Switzerland}
\address[uc]{Physics Department, University of Connecticut, Storrs, CT
  06269-3046, USA}
\address[bnl]{RIKEN-BNL Research Center, Brookhaven National
  Laboratory, Upton, NY 11973, USA}

\address[inr]{Institute for Nuclear Research of the Russian Academy of
  Sciences,
  % \\
  % 60th October Anniversary prospect 7a, 
  Moscow 117312, Russia}
\address[mipt]{Moscow Institute of Physics and Technology, 
  % MIPT, Institutsky per. 9, 
  Dolgoprudny 141700, Russia}

%\date{}

%%%%%%%%%%%%%%%%%%%%%%%%%%%%%%%%%%%%%%%%%%%%%%%%%%%%%%%%%%%%%%%%
\begin{abstract}
  We study the mechanism of generation of 7\keV{} sterile neutrino Dark
  Matter (DM) in the model with light inflaton $\chi$, which
  serves as a messenger of scale invariance breaking.  In this model
  the inflaton, in addition to providing reheating to the Standard
  Model (SM) particles, decays directly into sterile neutrinos. The
  latter are responsible for the active neutrino oscillations via
  seesaw type I mechanism.  While the two sterile neutrinos may also
  produce the lepton asymmetry in the primordial plasma and hence
  explain the baryon asymmetry of the Universe, the third one being
  the lightest may be of 7 keV and serve as DM. For this mechanism to
  work, the mass of the inflaton is bound to be light ($0.1-1$\GeV)
  and uniquely determines its properties, which allows to test the
  model.  For particle physics experiments these are: inflaton
  lifetime ($10^{-5}-10^{-12}$\,s), branching ratio of B-meson to
  kaon and inflaton ($10^{-6}-10^{-4}$) and inflaton branching ratios
  into light SM particles like it would be for the SM Higgs boson of
  the same mass.  For cosmological experiments these are: spectral
  index of scalar perturbations ($n_s\simeq0.957-0.967$), and amount
  of tensor perturbations produced at inflation (tensor-to-salar ratio
  $r\simeq0.15-0.005$).
\end{abstract}
%%%%%%%%%%%%%%%%%%%%%%%%%%%%%%%%%%%%%%%%%%%%%%%%%%%%%%%%%%%%%%%%

\maketitle

%%%%%%%%%%%%%%%%%%%%%%%%%%%%%%%%%%%%%%%%%%%%%%%%%%%%%%%%%%%%%%%%%%%%%%%%
\section{Introduction}

Discovery of the neutral scalar with properties very close to what we
expect for the SM Higgs boson \cite{Aad:2012tfa,Chatrchyan:2012ufa}
and absence of any definite hints of supersymmetry at LHC asks for its
replacement as a solution to gauge hierarchy problem.  Some hope is
associated with conformal or scale invariance that might be a symmetry
of the SM at tree level, but for the only dimensionfull parameter of
the SM $v$ which gives the vacuum expectation value to the
Englert--Brout--Higgs (EBH) field.  Yet this parameter may be
generated by the vacuum expectation value of a new scalar field (scale
invariance breaking messenger) introduced into particle physics, so
that the SM sector is scale invariant at tree level.

The field itself may be used to solve other SM problems.  Here we
discuss the idea that it may serve as an inflaton in the early
Universe.  The renormalizable model realizing this idea was suggested
in \cite{Shaposhnikov:2006xi} and further developed in
\cite{Anisimov:2008qs,Bezrukov:2009yw,Bezrukov:2013fca}.
In this Letter we assume that the SM sector of the model is scale
invariant, in order to alleviate the hierarchy problem of the Higgs
mass.  The only violation of scale invariance is assumed to be present
in the inflaton, which can be considered as a messenger of the scale
symmetry breaking, exact mechanism of this breaking is beyond the
present analysis.  Technically this means that we assumed that other
dimensionful parameters, like Higgs boson mass term or cubic terms in
the potential are small and can be neglected.
Introducing these terms
at electroweak scale would not change the phenomenology of
the model.
With only
one dimensionfull parameter explicitly breaking scale invariance in
the inflaton sector, the model is consistent with cosmological
observations \cite{Bezrukov:2013fca} and constraints from particle
physics related to the possible manifestation of the light inflaton in
B-meson decays \cite{Bezrukov:2009yw}.

The model may be further extended by introducing three Majorana
fermions $N_I$, $I=1,2,3$, which are singlets with respect to the SM
gauge group.  Yukawa-type coupling to inflaton provides these fermions
with Majorana mass terms when the inflaton field obtains vacuum
expectation value.  The Yukawa-type couplings between $N_I$, EBH
doublet, and SM lepton doublets lead to Dirac masses for neutrinos,
and the active neutrino masses are then obtained from seesaw type I
formula \cite{Minkowski:1977sc}.  Hence the fermions serve as sterile
neutrinos, and the Yukawa couplings in a part of the parameter space
may explain the baryon asymmetry of the Universe via leptogenesis,
e.g.\ implementing the $\nu$MSM scheme
\cite{Asaka:2005pn,Boyarsky:2009ix}.  Remarkably, the lightest sterile
neutrino $N_1$, provided tiny coupling to active neutrinos, may serve
as non-thermal DM produced by inflaton decays in the early Universe
\cite{Shaposhnikov:2006xi}.  Therefore the suggested model with seven
new degrees of freedom added to the SM explains the neutrino
oscillations, DM phenomena, baryon asymmetry of the Universe and
exhibits the inflationary dynamics at early times thus solving the Hot
Big Bang theory problems.

Given the allowed range of the inflaton mass the DM sterile neutrino
is naturally light here, $1\keV<M_1<1\MeV$.  In this \emph{Letter} we
discuss the particular choice of $M_1=7\keV$ motivated by recently
found anomalous line in cosmic X-ray spectra of galaxy clusters and
Andromeda galaxy observed by orbital telescopes
\cite{Abazajian:2001vt,Bulbul:2014sua,Boyarsky:2014jta}.  We outline the viable region
of the model parameter space consistent with this choice of sterile
neutrino mass and give definite predictions for the inflationary
cosmological parameters and the inflaton mass, its lifetime and
branching ratio of B-meson to inflaton which {\em allow to thoroughly
investigate this model}.  Remarkably, recent results of BICEP2
experiment \cite{Ade:2014xna} on detection of B-mode polarization,
interpreted as primordial tensor perturbations, can completely fix all
the parameters of the model.

The action of the light inflaton model augmented with three sterile
neutrinos is \cite{Bezrukov:2009yw,Bezrukov:2013fca}
\begin{align}
  S_{X\mathrm{SM}} =      &
    \int\!\!\sqrt{-g}\,d^4x\left( \cL_\mathrm{SM} + \cL_{X H}
      + \cL_{N} + \cL_\text{grav} \right)
    , \nonumber \\
  \cL_{XH} =              &
    \frac{(\partial_\mu X)^2}{2} + \frac{m_X^2 X^2}{2}
    - \frac{\beta X^4}{4}
    - \lambda \left( H^\dagger H - \frac{\alpha}{\lambda} X^2 \right)^2
    , \label{4*} \\
  \cL_\text{grav} =       &
    - \frac{M_P^2+\xi X^2}{2} R
    , \label{44} \\
  \cL_{N} = &
    i\bar N_I\slashed\partial N_I 
    - \left( F_{\alpha I}\bar L_\alpha N_I\tilde H 
      +\frac{f_I}{2}\bar N_I^c N_I X+\mathrm{h.c.} \right)
    , \label{4+}
\end{align}
where $R$ is the scalar curvature, $\cL_\text{SM}$ is the SM
Lagrangian without the EBH field potential, and
$\cL_{N}$ stands for the renormalizable extension 
of the SM by 3 sterile neutrinos $N_I$ ($I=1,2,3$), $L_\alpha$
($\alpha=e,\mu,\tau$) being lepton doublets and $\tilde H=\epsilon H^*$,
where $\epsilon$ is $2\times2$ antisymmetric matrix and $H$ is the EBH
doublet.

\begin{figure}[!htb]
  \centering
  \includegraphics{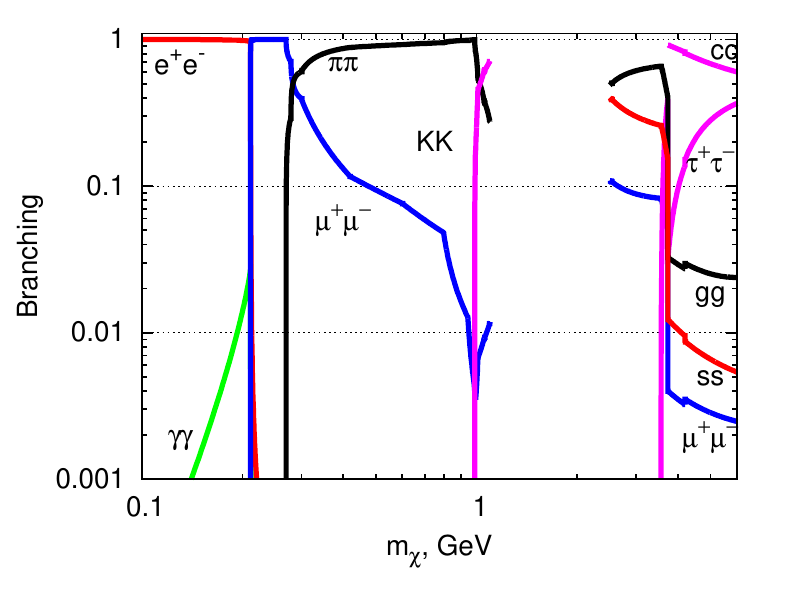}
  \caption{Inflaton decay branching rates \cite{Bezrukov:2009yw} 
    for two-body final states.  In the mass region $m_\chi\simeq 1\GeV$
    predictions are highly uncertain because of the QCD effects.}
  \label{fig:decays}
\end{figure}

With potential \eqref{4*} inflaton field $X$ gets vacuum expectation
value, which breaks scale invariance both in the sterile neutrino
sector (making sterile and active neutrino massive via \eqref{4+}) and
in the SM sector (giving vacuum expectation $v$ to the EBH field via
mixing in the last term of \eqref{4*}).  Four parameters of the model,
$m_X$, $\beta$, $\lambda$, and $\alpha$, determine the EBH field
vacuum expectation value $v\approx246\GeV$, the Higgs boson mass
$m_h\approx126\GeV$ \cite{Aad:2012tfa,Chatrchyan:2012ufa}, and the
inflaton mass
\begin{equation}
  \label{eq:10}
  m_\chi = m_h \sqrt{\frac{\beta}{2\alpha}}
         = \sqrt{\frac{\beta}{\lambda\theta^2}}
  .
\end{equation}
Thus, at a given value of $\beta$, the only free parameter in the
scalar sector is the mixing coupling $\alpha$ or the inflaton mass
$m_\chi$.  The particle spectrum in vacuum consists of the Higgs boson
$h$ and the inflaton $\chi$ of the mass $m_\chi$, which are mixed (as
compared to the $H-X$ basis) by a small mixing angle
\begin{equation}
  \label{eq:13}
  \theta^2 = \frac{2\beta v^2}{m_\chi^2} = \frac{2\alpha}{\lambda}
  .
\end{equation}
Hence, the branching ratios of the inflaton decay into the SM
particles (see \cite{Bezrukov:2009yw} for details) are fixed for a
given inflaton mass, see Fig.~\ref{fig:decays}.

At large field values potential \eqref{4*} exhibits slow roll
behavior along the direction $H^\dagger H=\alpha X^2/\lambda$, and
supports the inflationary expansion of the early Universe.  The
non-minimal coupling to gravity \eqref{44} allows to control the
amount of gravity waves generated at inflation \cite{Tsujikawa:2004my}
and for $\xi\gtrsim10^{-3}$ is consistent \cite{Bezrukov:2013fca} with
the Planck bounds \cite{Ade:2013uln}.  The tilt of the scalar
perturbation power spectrum also agrees with cosmological data.  For
a given $\xi$ the inflaton self-coupling $\beta$ is determined from the
amplitude of the primordial density perturbations
\cite{Bezrukov:2013fca}.

To summarize, the model (\ref{4*},\ref{44}) has three new parameters
$\xi$, $\beta$, and $m_\chi$ (or, equivalently, $\alpha$), in addition
to SM \footnote{Parameter $m_X$ can be traded for the SM
  parameter $v$}, and they are determined from the
following effects.
\begin{itemize}
\item $\beta$ and $\xi$ are related from the CMB normalization.
\item $m_\chi$ and $\beta$ are related by the requirement of the
  generation of proper abundance of DM (given DM mass $M_1$ or
  coupling $f_1$ is known).
\item $\xi$ can be determined from the measurement of the
  tensor-to-scalar ratio $r$ of the primordial perturbations.
\end{itemize}
This in principle completely fixes all the parameters of the model.
We treat $m_\chi$ as free parameter in this \emph{Letter}, as
far as the errors for the $r$ determination are still quite large, and
will discuss this further in the Conclusions.

At the same time the resulting values of the parameters should satisfy
the set of constraints \cite{Bezrukov:2009yw,Bezrukov:2013fca}
\begin{itemize}
\item $\alpha$ is bound from below from the requirement of sufficient
  reheating,
\item $\alpha$ is bound from above not to spoil the inflationary
  potential by radiative corrections,
\item certain region in $m_\chi$ and $\theta$ (or, equivalently
  $\beta$) is constrained from particle physics experiments.
\end{itemize}
We show below that the first two are automatically satisfied with the
parameters, leading to the proper DM generation, and the latter one
leads to significant bound on the inflaton mass $m_\chi$ (and hence effective
upper bound on $r$).

\section{Dark Matter generation}

The lightest sterile neutrinos in \eqref{4+} may serve as DM provided
its tiny mixing to active neutrinos keeps it sufficiently long-lived.
The dark matter particles may be produced in the primordial plasma via
inflaton decays \cite{Shaposhnikov:2006xi} due to Yukawa couplings in
\eqref{4+}.  They never come to equilibrium.

Let's discuss the production in details.  The light inflaton is in
thermal equilibrium down to rather small temperatures $T\ll m_\chi$,
thanks to reactions $\chi\leftrightarrow e^+e^-,\mu^+\mu^-$, etc.
Sterile neutrinos are produced in the inflaton decays mainly at
$T\simeq m_\chi$, and their distribution function $n(p,t)$ ($p$ is the
neutrino 3-momentum and $t$ is time) can be found from the solution of
the kinetic equation
\begin{equation}\label{kineq}
  \frac{\partial n}{\partial t} - \cH p\frac{\partial n}{\partial p}
  = \frac{2m_\chi\GNN}{p^2} \int_{p+m_\chi^2/4p}^\infty n_\chi(E) dE ,
\end{equation}
where the inverse decays $N_1N_1\to\chi$ are neglected, $\cH$ is the
Hubble constant, $E$ is the inflaton energy, $n_\chi(E)$ is the
inflaton (thermal equilibrium) distribution, $\GNN=\beta M_1^2/(8\pi
m_\chi)$ is the partial inflaton width for the $\chi\to N_1N_1$ decay.

One can obtain the analytic solution of (\ref{kineq}) in the
approximation of the time-independent effective number of degrees of
freedom \cite{Shaposhnikov:2006xi}, but this approximation is not
accurate enough for the interesting mass region $0.3\GeV\lesssim
m_\chi\lesssim1\GeV$.  Thus the numerical integration of (\ref{kineq})
is required together with the input of the hadronic equation of state
at GeV temperatures, which is not known exactly.  For the estimate one
can utilize the phenomenological equation of state from
\cite{Asaka:2006nq}.

\begin{figure}[!htb]
  \centering
  \includegraphics{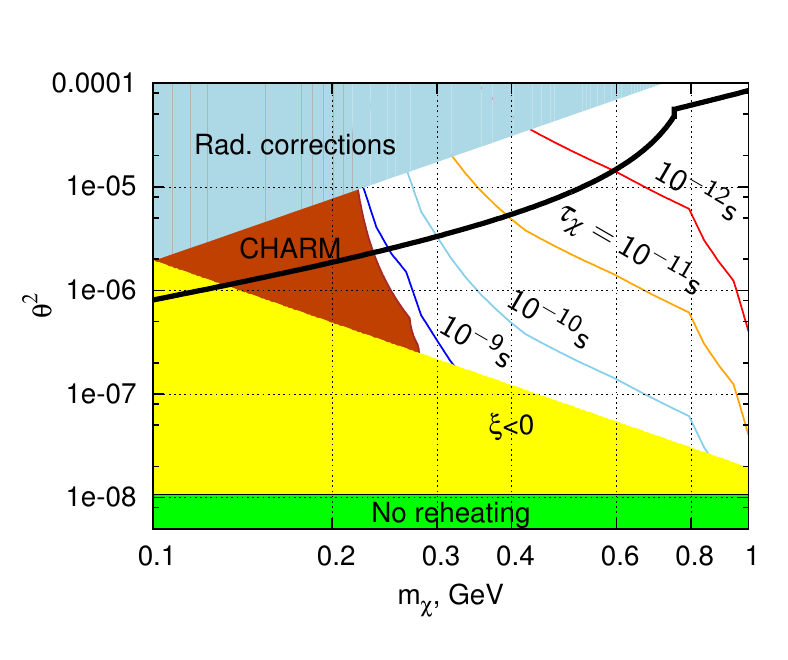}
  \caption{Higgs-inflaton squared mixing $\theta^2$ (thick black line)
    depending on the light inflaton mass $m_\chi$ for the model with
    sterile neutrino dark matter of 7\keV{} mass.  Various forbidden
    regions are shaded, and contours of the constant lifetime of the
    inflaton are shown. See \cite{Bezrukov:2013fca} for details.}
  \label{fig:particle-parameters}
\end{figure}

\begin{figure}[!htb]
  \centering
  \includegraphics{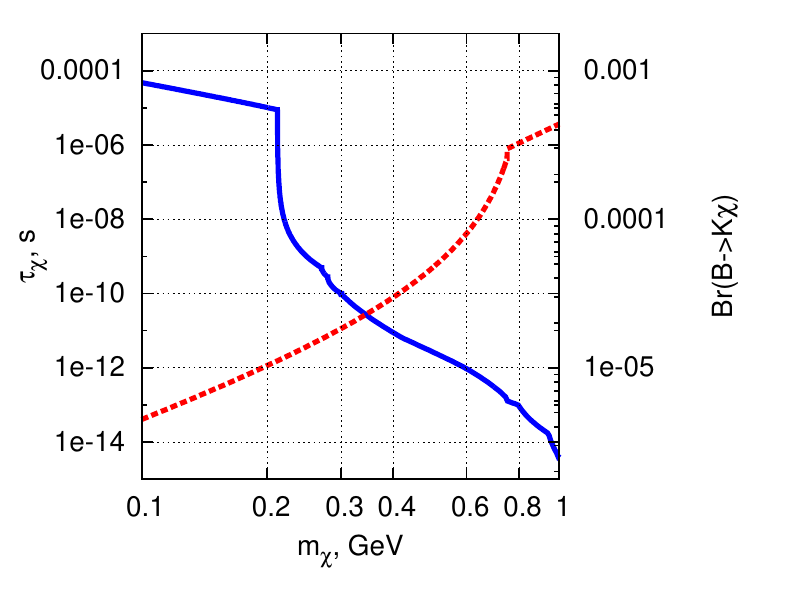}
  \caption{Inflaton lifetime (blue solid line, left vertical axis) and
    B-meson decay branching rate into inflaton and kaon (red dashed
    line, right vertical axis) as functions of the inflaton mass
    $m_\chi$ for the model with sterile neutrino DM of
    7\keV{} mass.}
  \label{fig:particle-features}
\end{figure}

Such integration was performed in \cite{Shaposhnikov:2006xi}, and here
we make use of these results.  The relative contribution of sterile
neutrino $N_1$ to the present Universe energy density $\Omega_N$ (must
be 0.25 to fully explain DM) 
determines at given $m_\chi$ and $M_1$ the value of 
the mixing angle (or quartic coupling $\beta$ by (\ref{eq:13}))
\begin{equation}\label{beta-for-mchi}
%   \beta \simeq 1.5\times10^{-10}
%   \left( \frac{7\keV}{M_1} \right)^3
%   \left( \frac{S}{6.4\, f(m_\chi)} \right)
%   \left( \frac{\Omega_N}{0.25} \right)
%   \left( \frac{m_\chi}{700\MeV} \right)^3, 
  % \beta \simeq 1.4\times10^{-12}
  % \left( \frac{7\keV}{M_1} \right)^3
  % \left( \frac{S}{1.5\,f(m_\chi)} \right)
  % \left( \frac{\Omega_N}{0.22} \right)
  % \left( \frac{m_\chi}{250\MeV} \right)^3, 
  \theta^2 \simeq 2.7\times10^{-6}
  \left( \frac{7\keV}{M_1} \right)^3
  \left( \frac{S}{1.5\,f(m_\chi)} \right)
  \left( \frac{\Omega_N}{0.22} \right)
  \left( \frac{m_\chi}{250\MeV} \right), 
\end{equation}
where $S\gtrsim1$ is a dilution factor accounting for a possible
entropy production due to late decay of the heavier sterile neutrinos
$N_{2,3}$~\cite{Asaka:2006ek} and the function $f(m_\chi)$ is
determined by the effective number of degrees of freedom $g_*(T)$ in
the primordial plasma at the inflaton decay.  It changes monotonically
from $0.9$ to $0.4$ for inflaton mass from 70\MeV{} to 500\MeV{} and
for heavier inflaton can be approximated as $f(m_\chi)\simeq \left[
  10.75/g_*(m_\chi/3)\right]^{3/2}$ ($g_*(T)$ can be obtained from
\cite{Asaka:2006nq}).  We further neglect possible dilution and
possible contribution of other dark matter production mechanisms
(e.g.\ neutrino oscillations amplified by lepton asymmetry in plasma
\cite{Shi:1998km}), which can increase or decrease $\beta$
respectively.  The mixing angle for $M_1=7\keV$ is plotted as a thick
line on
Figure~\ref{fig:particle-parameters}.  The case of different $M_1$ can
be easily reconstructed by shifting the line according to
(\ref{beta-for-mchi}).
The mixing angle
can be translated to the lifetime of the inflaton, see
Figure~\ref{fig:particle-features}.

\begin{figure}[!htb]
  \centering
  \includegraphics{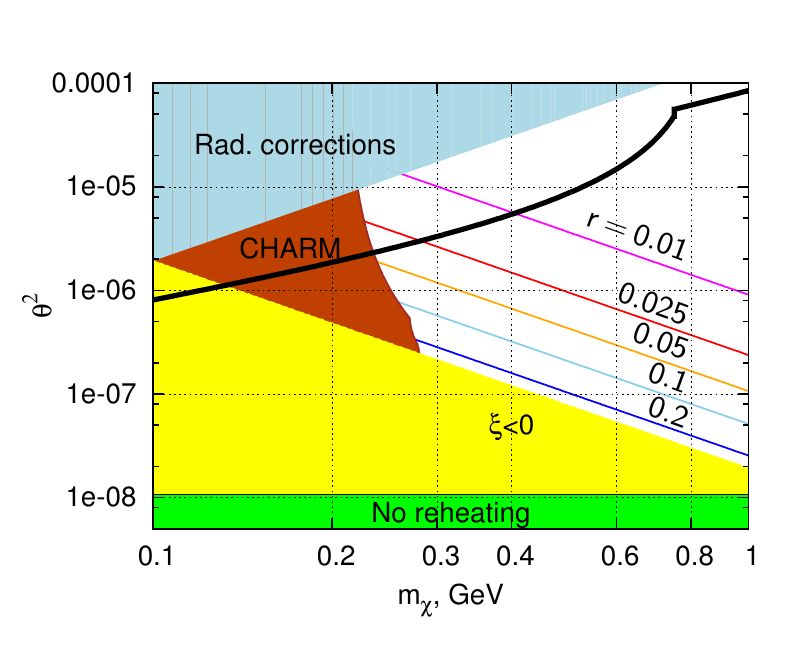}
  \caption{The same as Fig.~\ref{fig:particle-parameters}, but with
    lines of constant $r$.}
  \label{fig:particle-parameters2}
\end{figure}

The moment mixing angle is known the rates of meson decays to the
inflaton are determined.  The most promising for the inflaton searches
is the two-body decay of B-meson to kaon and inflaton, whose branching
ratio is \cite{Bezrukov:2009yw,Bezrukov:2013fca}
\begin{equation}
  \label{B-branching}
  \text{Br}(B\to\chi K)
%  &\simeq 10^{-6}\times \l 1-
%  \frac{m_\chi^2}{m_b^2}\r^2 \l \frac{\beta(\xi)}{1.5\times 10^{-13}}\r 
%  \l \frac{300\MeV}{m_\chi}\r^2 \\
%  &
  \simeq 4.8\times10^{-6} \times
  \l 1-\frac{m_\chi^2}{m_b^2} \r^2 \l \frac{\theta^2}{10^{-6}} \r
  ,
%  \notag
\end{equation}
and is also presented in Figure~\ref{fig:particle-features} for the
mixing $\theta^2$ corresponding to the proper DM generation
(Figure~\ref{fig:particle-parameters}).  Present accuracy in
measurements of (limits on) three-body decays of B-meson into kaon and
lepton pair, kaon and pion pair, kaon and kaon pair are at the level
of $10^{-5}-10^{-7}$ \cite{Beringer:1900zz} and they are certainly
relevant for short-lived inflaton, $\tau_\chi<10^{-12}$\,s.  At longer
lifetimes (lower masses), though the branching is still be rather
large (Figure~\ref{fig:particle-features}), the inflaton either
escapes the detector or gives rise to an event with a displaced decay
vertex.  Thus, the existing limits on three body B meson decays can
not be straightforwardly translated to the bounds on the inflaton and
additional analysis is required to find the exact region of parameters
excluded from B decay experiments.
%   Thus we
% conclude that for 7\keV{} DM the inflaton heavier than 1\GeV{} is most
% probably may be excluded from the analysis of these data.

The most significant experimental bound today can be obtained from the
CHARM experiment \cite{Bergsma:1985qz}.  In this experiment the
inflaton is created in the beam target and then the decay products of
the inflaton is searched in the detector which is placed at some
distance.  This method effectively bounds long lived inflatons
\cite{Bezrukov:2009yw}.  The corresponding bound is presented in
\ref{fig:particle-parameters}.

One can also check, that the required Higgs-inflaton coupling
$\alpha=\lambda\theta^2/2$ is within the bounds from reheating (to
happen in the early Universe before Electroweak sphaleron 
processes terminated) and
radiative corrections (not to spoil the inflaton potential)
\cite{Anisimov:2008qs} for all interesting inflaton masses, see
Figure~\ref{fig:particle-parameters}.

\begin{figure}
  \centering
  \includegraphics{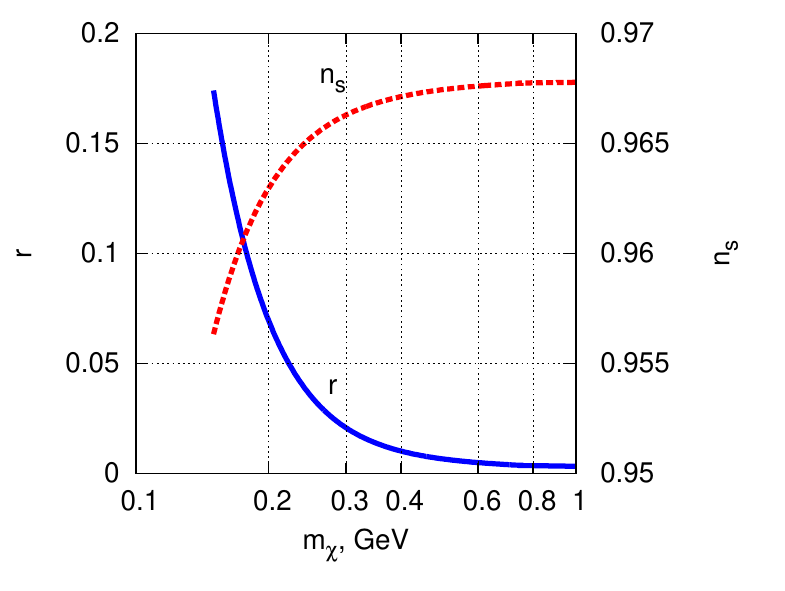}
  \caption{Predictions for the spectral index $n_s$ (red dashed line,
    right vertical axis) and tensor-to-scalar ration $r$ (blue solid
    line, left vertical axis) of the primordial density perturbations
    depending on the inflaton mass (assuming $M_1=7\keV$ DM
    production).}
  \label{fig:cosmo-parameters}
\end{figure}

The non-minimal coupling $\xi$ determines uniquely the spectral index
$n_s$ and the tensor-to-scalar ratio $r$ of the primordial
perturbations.  We give the cosmological predictions for the
interesting inflaton mass range in Figure~\ref{fig:cosmo-parameters}.
Note that \emph{measurement of the tensor modes $r$ fixes the value of
  inflaton mass.}

\begin{figure}[!htb]
  \centering
  \includegraphics{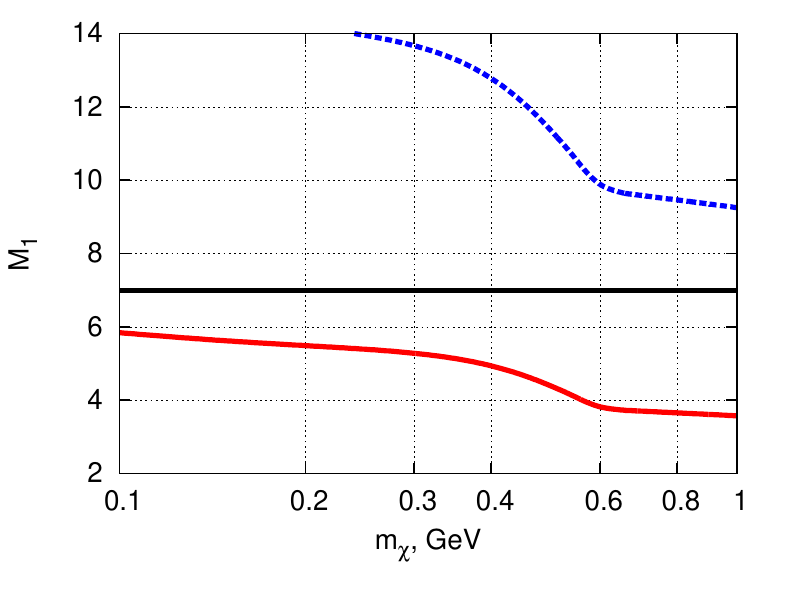}
  \caption{The Lyman-$\alpha$ lower bound on DM mass for DM produced
    by decays of the inflaton of mass $m_\chi$.  The lower and upper
    plots correspond to the bounds from \cite{Boyarsky:2008xj} and
    \cite{Viel:2013fqw}, respectively.  The 7\keV{} value is
    marked for reference.}
  \label{Ly-a}
\end{figure}

Another important property is the momentum distribution of the
generated DM neutrino.  There are stringent bound on the free
streaming length of a potentially Warm DM candidate from the analysis
of the Lyman-$\alpha$ forest \cite{Boyarsky:2008xj,Viel:2013fqw}.
While the exact reanalysis of the bounds is complicated, as far as one
should take into account the non-thermal shape of the spectrum, an
estimate can be obtained by simple comparison of the average momentum
of the generated DM neutrino.  This is a good approximation, as far as
in our case the spectral distribution does not have sharp
resonant-like features.  The average momentum of the neutrino is
\cite{Shaposhnikov:2006xi} (at temperatures $T$ just above neutrino
freeze-out)
\begin{equation}\label{average-momentum}
  \langle p \rangle
  \simeq 2.45\,T\left(\frac{10.75}{g_*(m_\chi/3)}\right)^{1/3},
\end{equation}
which is below the usual thermal average of $p_T=3.15\,T$.  One can
then deduce the mass bounds from the analysis of structure formation.
There are two types of bounds present in the literature: thermal relic
bounds $m_{TR}$ for the particles with the distribution of the thermal
shape but lower temperature; and non-resonantly produced neutrino
bound $m_{NRP}$, which has been obtained for $T=T_\nu$ and overall
suppressed distribution.  These bound can be translated to our case,
which is intermediate (both average momentum is below thermal and
overall distribution is suppressed), as
\begin{equation}\label{m-Lyalpha}
  m_{\text{Ly-}\alpha}(\langle p\rangle)
  = \frac{\langle p\rangle}{p_T} m_\text{NRP}
  = \frac{\langle p\rangle}{p_T} m_\text{TR}
     \left(
       \frac{m_\text{TR}}{(\Omega_\text{DM}h^2)94\eV}
     \right)^{1/3},
\end{equation}
where $h\approx 0.7$. 
For the inflaton mass in the interesting range the corresponding lower
limits on the fermion DM mass are outlined in Figure~\ref{Ly-a}.
The 7\keV{} DM is consistent with the first more conservative analysis
\cite{Boyarsky:2008xj} of Lyman-$\alpha$ forest data, but has tensions
with the later analysis.

\section{Summary}

In the model, where the non-minimally coupled inflaton serves as the
only scale invariance breaking messenger for SM with three
sterile neutrinos, sterile neutrino DM can be generated in the
inflaton decays\footnote{Two other sterile neutrinos giving masses to
  active neutrinos may be also adopted to explain baryon asymmetry of
  the Universe like in $\nu$MSM.}.  In bosonic sector the model introduces three
additional parameters---non-minimal coupling to gravity $\xi$, inflaton self
coupling $\beta$, and the inflaton mass $m_\chi$.  The amplitude of
the primordial perturbations relates first two of these parameters,
$\xi$ and $\beta$.  Requirement of the proper abundance of the DM with
a given mass $M_1$ provides the relation between the second pair,
$\beta$ and $m_\chi$.  Thus, assuming the mass of the DM is known, the
only free parameter left is the inflaton mass.  For numerical estimates
we take $M_1=7\keV$, motivated by recent results
\cite{Bulbul:2014sua,Boyarsky:2014jta}.  Further constraints on the
model can be made from inflationary observations, specifically the
tensor-to-scalar ratio $r$.  Exact knowledge of $r$ would fix the
value of $\xi$, and, thus, the inflaton mass $m_\chi$ leaving no free
parameters (see Figures~\ref{fig:particle-parameters2} and
\ref{fig:cosmo-parameters}).
 Recent
observation of $r\sim 0.2$ by BICEP2 \cite{Ade:2014xna} is in some tension
with the CHARM bound on our model.  However, if the dust foregrounds
give major contribution to the BICEP2 signal, $r$ may be smaller and
agree with predictions of our model.  Thus, improvement or verification
of the CHARM bound and independent checks of the BICEP2 result by
multifrequency experiments are essential.
Note also, that for heavier DM, $M_1>7\keV$, larger $r\sim0.2$ become allowed (cf.\
Figure~\ref{fig:particle-parameters2} and eq.\ (\ref{beta-for-mchi})).

The resulting low mass range is especially interesting, as far as for
the inflaton masses of $230\MeV\lesssim m_\chi\lesssim600\MeV$ the
inflaton can be produced and searched in B-meson decays, see decay
rate and inflaton lifetime in 
Figure~\ref{fig:particle-features}.  For the lower masses the lifetime
of the inflaton is relatively long and the most interesting signature
is the offset vertex of the inflaton decay into muon or pion pair
after the B-meson decay vertex.  For the higher masses the inflaton
lifetime drops rapidly, and the possible signature is the peak in the
B-meson three body decay kinematics.  Note that the expected event
rates are comparable with the current experimental sensitivity
\cite{Bezrukov:2013fca}. The lightest allowed inflatons may be
searched for in beam-dump experiments (e.g. see SHIP 
{\bf REF}
for CHARM successor). 

Finally, we should note, that there are ways to slightly relax the
relations for the model parameters.  First is the possibility of the
entropy generation in the decays of the heavier sterile neutrinos
after the DM generation ($S>1$).  This allows for slightly larger
$\beta$ for given $m_\chi$, leading to larger Higgs-inflaton mixing
$\theta^2$, larger $\xi$, and smaller $r$.  Additional generation of
DM sterile neutrino $N_1$ after the inflaton decay (as in
\cite{Boyarsky:2009ix}) leads to the opposite effect.  More
significant deviations is possible if we allow for additional sources
of the violations of the scale invariance.  Specifically, allowing for
arbitrary mass terms for the sterile neutrinos independent of the
inflaton coupling would allow to relax all the relations for the DM
generation.  If 7\keV{} sterile neutrino is not the dominant component
of DM, $\Omega_N<\Omega_{DM}$, Higgs-inflaton mixing $\theta^2$ is
smaller, hence larger $r$ is allowed.  It is worth to study thoroughly
the consistency of the particular mechanism of DM production with
observations of Ly-$\alpha$ forest, since present lower limits on the
DM free-streaming \cite{Boyarsky:2008xj,Viel:2013fqw} fall in the
right ballpark.

\bigskip

The authors would like to thank M.~Shaposhnikov for valuable
discussions.  The work of D.G. is partly 
supported by RFBR grants 13-02-01127a and 14-02-00894a.

%%%%%%%%%%%%%%%%%%%%%%%%%%%%%%%%%%%%%%%%%%%%%%%%%%%%%%%%%%%%%%
%\bibliography{Papers}
%\bibliographystyle{elsarticle-num}

\end{document}